\documentclass[12pt]{iopart}

\usepackage{graphicx}
\usepackage{epsfig}
\usepackage{txfonts}

\begin{document}

\title[Quadrant detection package]{A computational tool to characterize particle tracking measurements in optical tweezers}

\author{Michael~A.~Taylor and Warwick~P.~Bowen\footnote[1]{To whom correspondence should be addressed wbowen@physics.uq.edu.au}}

\address{Centre for Engineered Quantum Systems, University of Queensland, St Lucia, Queensland 4072, Australia}


\begin{abstract}
Here we present a computational tool for optical tweezers which calculates the particle tracking signal measured with a quadrant detector and the shot-noise limit to position resolution. The tool is a piece of Matlab code which functions within the freely available Optical Tweezers Toolbox. It allows the measurements performed in most optical tweezers experiments to be theoretically characterized in a fast and easy manner. The code supports particles with arbitrary size, any optical fields and any combination of objective and condenser, and performs a full vector calculation of the relevant fields. Example calculations are presented which show the tracking signals for different particles, and the shot noise limit to position sensitivity as a function of the effective condenser NA. 
\end{abstract}


\maketitle

\section*{Introduction}

 Optical tweezers have proven an indispensable tool for modern biophysics, and have advanced our understanding of a wide range of  single particle dynamic processes~\cite{Bustamante2004,Moffitt2008}. Consequently, the optical forces exerted in optical tweezers have been extensively studied and calculated~\cite{Harada1996,Rohrbach2002Trapping,Lock2004}. While it can be difficult and time-consuming to manually calculate the forces in a specific experiment, researchers can perform the difficult calculations with the freely available Optical Tweezers Toolbox~\cite{Nieminen2007,NieminenTB}. This determines the optical force for arbitrary incident fields and trapped particles, using a full vectorial field calculation, and allows researchers to rigorously model the trapping in their experiments in an easy and convenient manner.  The theory of particle tracking has also been well established for quadrant detection~\cite{Gittes1998,Rohrbach2002Position}. Additionally, the quantum limit to measurement sensitivity has been derived~\cite{Taylor2013QNL,Tay2009}, though this is only accessible with a perfect measurement which includes both the phase and amplitude of the light. Accurate calibration of an optical tweezers apparatus is important in almost all applications. Although particle tracking is well understood theoretically, currently no computational tool equivalent to the Optical Tweezers Toolbox is available to predict the response or sensitivity of such a system. Instead, researchers are required to repeat literature calculations for their specific apparatus. As a consequence, the use of theoretical tools to model detection is currently limited to a small subset of the community. The availability of a straightforward tool to perform such calculations would enable optimization of experiments and quantitative comparisons with theory.

Here we present a computational tool for optical tweezers which calculates the position signal measured with a quadrant detector, and the corresponding shot-noise limit to position sensitivity. This piece of Matlab code is designed to function within the Optical Tweezers Toolbox~\cite{Nieminen2007,NieminenTB}, and allows users to theoretically determine the measurement properties of their experiments without manually performing any calculations. Because the Optical Tweezers Toolbox can calculate the scattering of arbitrary optical fields from any trapped particle, this code inherits the same versatility. The source code is available at~\cite{OnlinePackage}.

The calculations performed by this code have several applications. For instance, in most optical tweezers measurements it is important to be able to determine the range over which the measurement is linear.  The code allows this range to be directly predicted, and through this provides a convenient tool to optimize the linearity by varying experimental parameters such as numerical aperture or particle size. By providing the signal amplitude retrievable from a given apparatus, the code also makes it possible to determine whether a given phenomena will be measurable, and to optimize the experimental apparatus to maximize signal-to-noise where required.

\section*{Principle}

\begin{figure}
 \begin{center}
   \includegraphics[width=8cm]{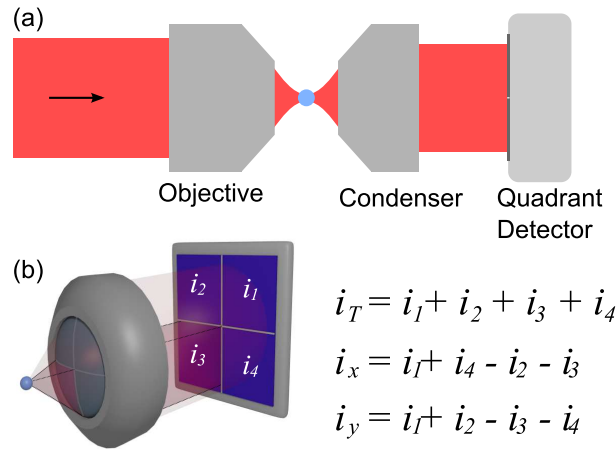}
   \caption{ This schematic describes the particle tracking setup which the code calculates. (a) The incident field is focused by the objective, and then interacts with a particle. The transmitted field is then collected by a condenser and directed onto a quadrant detector at the back focal plane. (b) The transmitted light which is in one quadrant at the condenser remains in this quadrant until it reaches the detector. The signal measured on the quadrant detector can be evaluated by calculating the intensity profile at the farfield of the particle, and integrating over the area within the condenser aperture in the relevant quadrants. The measurements are the two subtraction currents $i_x$ and $i_y$, and the total photocurrent $i_T$ which are defined as shown here.
}
 \label{QuadrantLayout}  
 \end{center}
\end{figure}

 In an optical tweezers experiment, an objective focuses an incident field to a spot, where particles are trapped by optical forces. This interaction includes a momentum exchange between the field and the particle, which therefore changes the propagation direction of the light. This has an overall effect of deflecting the transmitted field. To measure this deflection, the transmitted light is collected with a high numerical aperture (NA) condenser lens and measured at the back focal plane with a quadrant detector~\cite{Gittes1998} (See Fig.~\ref{QuadrantLayout}).  The condenser changes the spatial distribution of the light which passes through its aperture, without changing the photon flux arriving in each quadrant. Therefore, the signal measured on the quadrant detector can be evaluated by calculating the power within the four quadrants of the condenser aperture at the farfield of the particle~\cite{Rohrbach2002Position} (See Fig.~\ref{QuadrantLayout}(b)).
 
  Many real experiments do not place the detector at the back focal plane as shown in Fig.~\ref{QuadrantLayout}(a), but use an additional lens  to image the field at the back focal plane onto the detector. This additional lens also has a finite aperture width which introduces additional clipping,  effectively reducing the condenser NA. In order to accurately model the measurement, the effective condenser NA which includes any additional clipping must be used.  

\section*{Calculation of the signal}

 To use the supplied code, the user needs only to define the experimental parameters. The relevant parameters are the optical wavelength in vacuum, the refractive indices of the medium and particle, the particle radius (assuming spherical particles), the NA of the objective and condenser, the measured optical power, the optical polarization, and the spatial profile of the incident field in the Laguerre-Gaussian basis. Once these are defined, the code uses functions present in the Optical Tweezers Toolbox~\cite{Nieminen2007,NieminenTB} to decompose the incident field into an expansion of spherical harmonics, given by the coefficients $a$ and $b$. Then the T-matrix is calculated for the scattering particle~\cite{Nieminen2011}. The location of the axial trapping point is determined following the examples in the toolbox. To find the measured signal, the scattered field coefficients $p$ and $q$ are calculated with the particle at a range of transverse displacements from the trapping point.
 The coefficients $a$, $b$, $p$ and $q$ fully determine the transmitted optical field, and therefore allow calculation of the optical force (which the Optical Tweezers Toolbox was designed for) and the  intensity profile which is measured on the detector.  The particle tracking response can also be calculated for scatterers which are not homogeneous spheres, with the files ``Quadrant$\_$measurement$\_$layered.m'' and ``Quadrant$\_$measurement$\_$cube.m'' respectively calculating the particle tracking response for a layered sphere and a cube.

\begin{figure}
 \begin{center}
   \includegraphics[width=8.5cm]{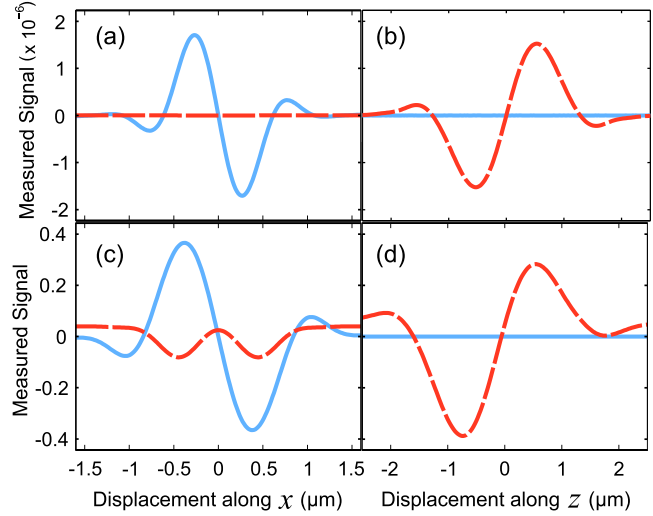}
   \caption{  The particle tracking signals photocurrents are shown for a measurement of a 10~nm (a, b) and a 1~$\mu$m (c, d) diameter polystyrene ($n=1.58$) sphere in water ($n=1.33$) as it moves transversely along the $x$ axis (a, c) and axially along the $z$ axis (b, d). The setup uses 1064~nm light polarized linearly along the $x$ axis, and with objective and condenser NA of 1.2 and 1.0 respectively. The subtraction signal $i_x / \langle i_T \rangle$ (solid blue line) provides effective tracking of particle displacements along the $x$ axis, where $\langle i_T \rangle$ is total photocurrent for the centered particle. The total collected light $\left( i_T-\langle i_T \rangle\right)/\langle i_T \rangle$ (dashed red line) provides information about the displacement along the $z$ axis, although this is also affected by transverse motion, particularly for large beads (c). }
 \label{Signal_dx}  
 \end{center}
\end{figure}

The optical field is then calculated over an angular grid of points to find the photocurrent in each quadrant. This calculation can be performed with the ``farfield'' function in the toolbox, although this is not efficient when running a sequence of calculations with the same particle. To reduce the computation time, we use a modified version of this function which calculates two matrices $A_p$ and $B_q$ which implicitly contain the angular grid. The transmitted electric field is determined by the matrix multiplications
\begin{equation}
E(\theta,\phi )= A_p(\theta, \phi ) \times \left( a+2p \right) + B_q(\theta, \phi ) \times \left( b+2q \right).
\end{equation}
Then, the light intensity is integrated to determine the photon flux present in each of the quadrant detector signals. The particle tracking signals $i_x$ and $i_y$ are determined by subtracting the light incident on one half from the other, as defined in Fig.~\ref{QuadrantLayout}(b). Additionally, the total photocurrent $i_T$ is calculated as this can be used to determine the particle position along the $z$ axis~\cite{Pralle1999,Dreyer2004}. These are calculated as photon numbers, so the total photocurrent $i_T = P/(\hbar \omega)$, where $P$ is the detected optical power and $\hbar \omega$ is the energy per photon. Since detectors in real experiments are not perfectly efficient, not every incident photon is measured. To account for this, the theoretical power $P$ should be lower than that used in experiments by a factor given by the detection efficiency. Examples of the calculated photocurrent signals are shown in Fig.~\ref{Signal_dx}, which agree well with previously published calculations~\cite{Gittes1998,Rohrbach2002Position,Tay2009}. 

\section*{Measurement sensitivity}

The sensitivity of most real experiments is limited by electronic noise in the detectors, though there are methods to improve this~\cite{Taylor2013LockIn,Chavez2008}. With sufficient improvements, the detection will eventually be shot-noise limited. At this point the sensitivity can no longer be improved by reducing the electronic noise, and further improvements require use of alternative strategies~\cite{Taylor2011,Taylor2013_sqz}. The code presented here determines this shot-noise limit from the calculated detector response to particle displacements, and thus allows the gap between the shot-noise limit and the experimentally achieved sensitivity to be quantified. This also provides a straightforward method to compare the strength of the detection signal that different particles or trapping configurations produce.

The subtraction photocurrent $i_x$ responds linearly for small displacements along $x$, so in this limit we can define a gain $G_x$ by
\begin{equation}
\langle i_x \rangle = \langle i_T \rangle G_x x,\label{i_x}
\end{equation}
where $\langle 	i_T \rangle$ is the mean total number of photons measured on the detector. If the light is in a coherent state, and there is no additional noise, then the photons follow Poissonian statistics and the quantum shot noise of this measured signal is given by 
\begin{equation}
\langle \Delta i_x^2 \rangle =\langle i_T \rangle.\label{Var_i}
\end{equation}
Combining Eq.~\ref{i_x} and Eq.~\ref{Var_i}, we find that the signal-to-noise ratio is given by 
\begin{equation}
 \frac{\langle i_x \rangle^2}{\langle  \Delta i_x^2 \rangle} = \langle i_T \rangle G_x^2 x^2.\label{SNR_x}
\end{equation}
 A displacement is resolvable when it yields a signal-to-noise ratio greater than 1, so the minimum resolvable displacement in units of m/Hz$^{-1/2}$ is 
\begin{equation}
x_{\rm min} = \langle i_T \rangle^{-1/2} G_x^{-1}.\label{x_min}
\end{equation}
The gain $G_x$ is defined by the calculated detector response, and this is used in the presented code to reveal the shot noise limit to position sensitivity. A similar calculation finds the shot noise limit along the $y$ and $z$ axes. For the data in Fig.~\ref{Signal_dx}(c) and (d), for instance, a quadrant measurement which perfectly captures 1~mW of optical power should have a displacement sensitivity  at the levels 9.3$\times 10^{-15}$~m/Hz$^{-1/2}$ and 1.7$\times 10^{-14}$~m/Hz$^{-1/2}$ along the $x$ and $z$ axes respectively. If this is measured with 1~$\mu$s time resolution, then the measurement bandwidth is 1~MHz and the minimum resolvable displacements along $x$ and $z$  are respectively 9.3~pm and 17~pm. Although such sensitivities are impressive, recent experiments have demonstrated a position sensitivity within a factor of 2 of the transverse prediction~\cite{Chavez2008}. It is important to note that both axial and transverse motion effect the sum photocurrent $i_T$ (see Fig.~\ref{Signal_dx}(c)), so the axial position cannot be accurately determined without accounting for a transverse displacement. Also, the sum photocurrent is sensitive to laser amplitude noise which cancels from the subtraction signals $i_x$ and $i_y$. This means that the axial position sensitivity is likely to be further from optimal in a real experiment than the corresponding transverse sensitivities.

\begin{figure}
 \begin{center}
   \includegraphics[width=8cm]{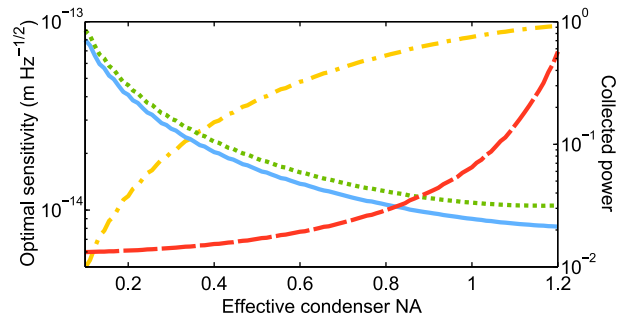}
   \caption{ The best displacement resolution possible with a quadrant detector, for a 1~$\mu$m  polystyrene sphere in water tracked with 1~mW of measured 1064~nm light polarized linearly along the $x$ axis, which is focused with an NA=1.2 objective. As the effective condenser NA increases, the sensitivity along the $x$ (solid blue line) and $y$ (green dotted line) axes improves. The axial sensitivity (red dashed line), however, improves with decreasing condenser NA.  The proportion of the trapping power which can be collected (yellow dash-dot) rises with increasing condenser NA, as this increases the angular range which can enter the detector.}
 \label{SensNA}  
 \end{center}
\end{figure}

The code allows the shot noise limit to sensitivity of particle tracking in all three axes to be quantified for particles of any size, trapped in arbitrary optical fields, and measured with any condenser NA. For instance, the sensitivity attainable is calculated as a function of effective condenser NA,  for a fixed objective (NA=1.2) and 1~mW of measured optical power (shown in Fig.~\ref{SensNA}). In this case, the position sensitivity along the transverse directions improve with NA, while the axial sensitivity per measured photon degrades, as demonstrated in Refs.~\cite{Rohrbach2002Position,Dreyer2004}. As such, the optimal method for three-dimensional particle tracking is to use a quadrant with a large capture angle to monitor the transverse position, while a second bulk detector with a small capture angle monitors the axial position.

\section*{Conclusion}

 We have presented a piece of Matlab code which works within the Optical Tweezers Toolbox to calculate the response of a quadrant detector to particle displacements, and the associated shot-noise limit to displacement sensitivity. The code is available at~\cite{OnlinePackage}. The calculation supports particles with arbitrary size, any optical fields and any combination of objective and condenser.

\section*{Acknowledgments}

We would like to thank Alexander Stilgoe and Timo Nieminen for advice about the Optical Tweezers Toolbox. This work was supported by the Australian Research Council Discovery Project DP0985078 and Centre of Excellence for Engineered Quantum Systems CE110001013.

\section*{References}

\bibliographystyle{jphysicsB}

\end{document}